\documentstyle [prd,preprint,aps] {revtex}
\begin{document}
\tighten
\draft
\preprint{IFP-472-UNC}
\title{What Can Be Learned with an Iodine Solar--Neutrino Detector?}
\author{J. Engel}
\address{Department of Physics and Astronomy, CB3255, University
of North Carolina,\\ Chapel Hill, North Carolina 27599}
\author{P. I. Krastev\cite{byline}}
\address{Institute for Advanced Study, Princeton, New Jersey 08540}
\author{K. Lande}
\address{Department of Physics and Astronomy, University of
Pennsylvania,\\
Philadelphia, Pennsylvania 19104}
\maketitle
\begin{abstract}
We study the potential benefits of an iodine-based solar--neutrino
detector for testing hypotheses that involve neutrino oscillations.
We argue that such a detector will have a good chance of distinguishing the
two allowed regions of $\Delta m^2$ -- $\sin^22\theta$ parameter space
if neutrino conversion is occurring in the sun.  It should also be
able to detect seasonal variations in the signal due to vacuum
oscillations and might be sensitive enough to detect day/night
variations due to MSW transitions in the earth.  Although it would need to be
calibrated, a working iodine
detector could be completed before more ambitious projects that
seek to accomplish the same things.
\end{abstract}

\narrowtext

Our current understanding of solar neutrinos comes either from
water--Cherenkov or radiochemical detectors.  Water--Cherenkov
detectors have several advantages over their radiochemical
counterparts: they are real-time detectors, often with a high count
rate, and can provide information on the direction of the incident
neutrinos.  The existing Cherenkov detectors, however, can measure
only the flux of high--energy neutrinos from the decay of $^8$B.
Radiochemical detectors, on the other hand, are unable to determine
either the direction of the neutrinos or the exact time of capture,
but {\em can} measure the flux of lower energy neutrinos.  Because
they rely only on charged-current interactions, they are also better
positioned to test neutrino--flavor--oscillation hypotheses; in
existing water--Cherenkov detectors neutral current neutrino--electron
scattering tends to wash out the variations in signal strength that
might result from conversion of solar electron--neutrinos ($\nu_e$)
into other neutrino flavors ($\nu_{\mu}$ or $\nu_{\tau})$.

It was pointed out some time ago\cite{hx1} that $^{127}$I, operating
via the reaction
\begin{equation}
\nu_{\rm e} + \hskip 2pt
^{127}{\rm I}\rightarrow {\rm e}^- + \hskip 2pt ^{127}{\rm Xe}~~,
\end{equation}
would make a useful solar--neutrino detector.  The effective threshold
for the reaction (1) is 0.789 MeV, low enough to enable the detection
of $^7$Be-, pep-, CNO- and $^8$B--neutrinos.  The cross sections for
these neutrinos in iodine were recently computed in Ref.\ \cite{epv}.
If that calculation is correct the total event rate in an iodine
detector should be 36 SNU, much larger than in chlorine (with the
standard--solar--model neutrino fluxes of Bahcall and Pinsonneault\cite{BP}).
What's more,
within the same assumptions, iodine is predicted to be particularly
sensitive to $^7$Be--neutrinos --- roughly 14 SNU from that source
alone are expected.

Here we show that if these predictions are close to the truth, an
iodine detector could play an important role in resolving the
solar--neutrino problem. At present, solutions involving purely solar
physics are regarded as implausible because they usually require that
the flux of $^8$B--neutrinos be more suppressed more than that of the
other solar--neutrino types. The data, by contrast, indicate that
the $^7$Be--neutrino flux is most suppressed\cite{hatal,kwr,jb1,del,parke}.
The predicted sensitivity of iodine to
$^7$Be--neutrinos makes it particularly suited to test this result.
The same prediction also makes iodine useful for testing temporal
variations of the signal, due either to MSW oscillations of neutrinos
as they pass through the earth (producing a day/night
variation\cite{hatal2}) or to vacuum oscillations (producing a
seasonal variation associated with changes in the distance between the
earth and sun\cite{justso}).  Other beryllium--sensitive detectors are
under development but are years away from actual operation, while a
prototype iodine detector should be ready by sometime this summer.

We begin by examining the signal in an $^{127}$I detector if the MSW
effect\cite{MSW} --- the resonant conversion of solar $\nu_e$'s to
other neutrino species inside the sun --- is the solution to the
solar--neutrino problem.  Existing experiments have put stringent
constraints on the difference $\Delta m^2$ between the squares of the
masses of the neutrinos and on the mixing--angle $\theta$, the
two parameters that determine the transition probability
$P(\nu_e\rightarrow\nu_{\mu})$.  Only two regions in the space of
these parameters are still allowed: a small mixing-angle solution with
$\Delta m^2
\simeq 6\times10^{-6}$ eV$^2$ and $\sin^22\theta \simeq 5\times10^{-
3}$, and a large mixing--angle solution with $\Delta m^2 \simeq 1.5
\times 10^{-5}$ eV$^2$ and $\sin^22\theta \simeq 0.8$.  The quality of
the fit in the small mixing--angle region is considerably better than
in the large mixing--angle region\cite{KP}. The two regions, allowed with
95\% confidence,
are shaded in all of the graphs in Fig.~1.

To understand the progress that might be made with an iodine detector,
we computed expected event rates throughout the two--dimensional
parameter space.  The corresponding iso--SNU contours, representing
curves of constant event rate in the iodine detector, are shown in
Fig.~1b, where we have assumed that the neutrino capture cross sections are
exactly those calculated in Ref.\ \cite{epv}.  With no MSW suppression,
the total event rate in iodine is then 36.4 SNU (the contributions of
the individual neutrino types are 18.4, 14.0, 1.85, 0.727 and 2.43 SNU
for the $^8$B, $^7$Be, pep, $^{13}$N and $^{15}$O neutrinos
respectively).  The figure shows that in the small mixing--angle
region, the total rate is reduced to between 5 and 13 SNU, while in
the large--angle region it lies between 10 and 16 SNU .  It may
therefore be possible to distinguish between the two solutions if the
detector clearly registers either less than 10 SNU or more than 13
SNU.  The reason for the difference in count rates is that in the
small mixing--angle region the $^7$Be--neutrinos are converted almost
completely into other neutrino types, while the flux of the
higher--energy neutrinos is reduced only by about 50\%.  In the
large--angle region, by contrast, all fluxes are reduced by 60--70\%.
The predicted sensitivity to ${}^7$Be--neutrinos means that the count
rates in the two allowed regions will differ more from one another
than they do in existing detectors, which are less sensitive to these
neutrinos.

${}^{127}$I is a complicated nucleus, however, and so the
cross sections calculated in Ref.\ \cite{epv} carry significant
uncertainty.  The model used there was designed to represent an entire
spectrum of states and the strength to any particular one, e.g. the
state accessible to ${}^7$Be neutrinos, is highly uncertain.  The
results of several calculations, within two related but distinct
frameworks in Ref.\ \cite{epv} and in an entirely different model in
Ref.\ \cite{iach}, prompted us to assign a conservative (but tentative)
uncertainty of $\pm$~80\% to the ${}^7$Be--neutrino cross section.
The CNO- and pep- neutrinos can access several low--lying states and so
we assigned a nominal uncertainty of $\pm$~60\% for their absorption
rate (which is strongly correlated with the ${}^7$Be rate).  Obtaining
an uncertainty for the ${}^8$B--neutrino cross section was slightly
more complicated.  On the one hand, within any particular model, the
uncertainty is quite small because of the large number of states
involved.  Moreover, the results of Ref.\ \cite{epv} are in reasonable
agreement with the strength distribution inferred from (p,n)
measurements.  Unfortunately, the proportionality constant relating
the two distributions is at present impossible to determine in odd--A
nuclei.  Although a sum rule prevents the constant from straying too
much, the uncertainty is on the order of 50\%, a fact reflected in
Ref.\ \cite{epv} by the use of two distinct values for $g_A$, the effective
axial--vector coupling in nuclei.  Here we adopted the 50\% as a rough
measure of the uncertainty in our integrated ${}^8$B cross section.

What happens to the iso--SNU curves when the cross sections are varied
within the estimated limits?  To find out we recalculated event rates
for 11 sets of cross sections that differ from those used in Fig.~1b
by amounts within the uncertainty ranges just discussed.  We assumed
that the CNO- and pep--uncertainties are completely correlated with
that of $^7$Be.  The results are shown in Table~1 and the
corresponding iso--SNU curves for two extreme cases appear in Fig.~1a
and Fig.~1c.  The table shows that for all the sampled cross section
sets there is some chance of distinguishing the small and large
mixing--angle regions.  Furthermore, the chances are strongly
correlated with the ratio of $^7$Be to $^8$B cross sections; the
higher this ratio, in general, the more likely it is that the count
rate will be consistent with one region and inconsistent with the
other.  Unless the ratio is significantly lower than the value in
Ref.\ \cite{epv}, an iodine detector could well pin down the
mass and mixing angle of the electron neutrino.

The detector might even be able to test the MSW hypothesis in another
way.  If the hypothesis is correct, oscillations can occur inside the
earth as well as the sun and cause a day/night variation in a
detector.  For the effect to be measurable the ratio of $^7$Be to
$^8$B cross sections must be large enough so that variations of the
event rate due to oscillations of the ${}^7$Be--neutrinos can be
observed despite the ``background'' contribution from boron neutrinos.
Our results indicate that if the cross sections are indeed as high as
calculated in Ref.\ \cite{epv}, this task should be achievable.

We turn now to another scenario; despite all the attention paid to the
MSW hypothesis, neutrino oscillations in vacuum remain a possible
solution to the solar--neutrino problem, consistent with all existing
data\cite{KPvac}.  A distinctive feature of vacuum oscillations is a
seasonal variation in the event rate due to the eccentricity of the
Earth's orbit.  The variation is most pronounced for monoenergetic
neutrinos.  Detectors in which $^7$Be--neutrinos supply a
significant portion of the total signal are therefore better able to test
this hypothesis than are water--Cherenkov detectors.  Here, as in the case
of day/night oscillations, if the value for the $^7$Be--neutrino
cross section is close to the one computed in
Ref.\ \cite{epv} an iodine detector would be more useful than any of
the existing detectors.  In Fig.~2 we have plotted the predicted
seasonal variations in iodine, chlorine, and gallium detectors for six
pairs of parameters $\Delta m^2$ and $\sin^22\theta$ that are still
allowed by the solar--neutrino data. The normalization $R$ of the signal,
described in detail in Ref.\ \cite{krpe}, is obtained by dividing the signal
measured in a particular run by the average observed over one year.
Presented in this form the variations of the signal do not depend on
the total neutrino flux from the Sun, but only on the parameters
$\Delta m^2$ and $\sin^22\theta$ and on the individual detector
characteristics.  As is clear from Fig.~2, the amplitude of the
seasonal variations in all six cases is considerably larger in the
iodine than in the other two detectors.

Before an iodine detector can truly be useful, of course, it will have
to be calibrated at least to some degree; the arguments presented here
rely on calculations that as we have noted carry considerable
uncertainty.  Fortunately, a program to directly measure the
cross section of neutrinos on $^{127}$I at different energies is
underway.  The program consists of three parts: a measurement of the
total cross section for stopped--muon--decay neutrinos at LAMPF, a
more direct measurement of the cross section for $^8$B--neutrinos,
and the use of an intense $^{37}$Ar source to measure the
cross section for $^7$Be--neutrinos.

The LAMPF measurement (already underway) is intended to determine the
sensitivity of an iodine detector to $^8$B--neutrinos from its
response to a beam of neutrinos produced by the decay of stopped
muons.  Located 7 meters from the proton beam stop and heavily
shielded, the detector is filled with 1.5 tons of iodine ---
$7.5\times10^{27}$ atoms of $^{127}$I --- in the form of sodium iodide
dissolved in water.  $^{127}$Xe is periodically extracted from the
detector; preliminary measurements were recently
reported in Ref.\ \cite{crconf}.  Ref.\ \cite{epv} argues that it will
probably be difficult to extract {\em solar}--neutrino cross sections from
this experiment, but that conclusion may change if calculations of
forbidden transitions improve in accuracy or if the LAMPF
cross section turns out to be smaller than the preliminary result.
In any event the analysis is continuing.

More direct measurements of the response of $^{127}$I to
$^8$B--neutrinos are possible.  The reaction $^6{\rm Li}(^3{\rm
He,n})^8$B can be used to produce an intense source of $^8$B; the
cross section for the production reaction was measured some time ago
by Marrs et al.\cite{marrs}. Another approach is to determine the
neutrino cross section as a function of energy by measuring the energy
of each neutrino from a LAMPF--like source that interacts in a
NaI--crystal electronic detector.  The secondary--particle interaction
signal, coming from the outgoing electron and gammas observed in the
NaI crystal, would fix the energy of the incident neutrino.  Since the
neutrino spectrum from the decay of stopped muons is well known,
the corresponding cross section could then be determined.  The
construction of an appropriate detector is currently
under consideration.

The most important quantity for our purposes is clearly the response to
$^7$Be--neutrinos.  The ideal calibration source is
$^{37}$Ar\cite{hx2}, which gives rise to monoenergetic neutrinos of
0.814 MeV. $^{37}$Ar can be produced via neutron capture by $^{36}$Ar
or via the reaction $^{40}{\rm Ca}(\rm{n},\alpha)^{37}$Ar.
Ref.\ \cite{gavrin1} reports preliminary plans to use the latter reaction to
produce a megacurie source of $^{37}$Ar.  The hope is that this
experiment will measure this crucial cross section with a precision of 10\%.

A 1/10 scale iodine solar--neutrino detector containing 100 tons of
$^{127}$I ($5\times10^{29}$ atoms) is now under construction at the
Homestake Mine.  A total interaction rate of 11.5 SNU, the dividing
rate between the small and large mixing--angle solutions in Fig. 1,
would lead to 0.5 $^{127}$Xe atoms per day in the 100 ton detector.
With corrections for decay before extraction and for counting
efficiency, this gives 120 observed $^{127}$Xe decays per year.  Even
with the prototype detector, a 20\% seasonal variation, due either to
neutrino oscillations in vacuum or to resonant conversion in the
earth, could probably be detected at the $3\sigma$ level with two years of
data.  The full--scale iodine detector could of course do much better.

Several other solar--neutrino detectors are planned for the next few
years; two of them --- SNO\cite{SNO} and
Superkamiokande\cite{superK} --- are expected to start taking data in
the near future. The Icarus experiment\cite{icar}, though not yet fully
funded, may also soon be on line.  But none of these three
detectors will be able to see $^7$Be--neutrinos.  Thus while they will
help us understand the solar--neutrino deficit, they are
unlikely to completely solve the problem.  The Borexino
experiment\cite{borex}, on the other hand, is designed precisely to measure
the $^7$Be--neutrino flux, but is still years away from operation;
futhermore, problems with background from natural radioactivity have
not yet been resolved.  Two helium experiments\cite{heron,helaz} have
ambitious plans to measure both the pp- and $^7$Be--neutrinos, but are
still in early stages of their development and seem unlikely to be
completed before the end of the century.  By contrast, a full--scale
iodine detector could be running in two or three years.  Provided it
can indeed be calibrated, an iodine detector therefore offers the best
hope for early resolution of the solar--neutrino problem.

This work was supported in part by the U.S. Department of Energy under
grant DE-FG05-94ER40827, by the National Science Foundation under
grant PHY93-12480, and by Dyson Visiting Professor Funds from the
Institute for Advanced Study.

\begin{figure}
\caption{Iso--SNU contours for an iodine detector
superimposed on the regions still allowed with 95\% confidence. Labels on the
curves are event rates
in SNU's. In Fig.~1b it has been assumed that the neutrino--capture
cross sections are those calculated in Ref.\ \protect{\cite{epv}}, while in
Fig.~1a (Fig.~1c) the integrated cross sections have been assumed to
be smaller (larger) by 50\% for the $^8$B--neutrinos, 60\% for the
CNO--neutrinos and 80\% for the $^7$Be--neutrinos.}
\end{figure}

\begin{figure}
\caption{
Annual variations $R$ (see text) of the signals in (a) chlorine, (b) iodine,
and (c) gallium detectors due to neutrino oscillations in vacuum. The values
of the parameters ($\Delta m^2$/eV$^2$, $\sin^22\theta$) corresponding
to each curve (all still allowed) are ($5.4\times 10^{-9}$, $1.0$) --- full
line; ($1.1\times
10^{-10}$, $0.95$) --- dotted line; ($9.1\times 10^{-11}$, $0.85$) ---
short-dashed line; ($8.1\times 10^{-11}$, $0.80$) --- long-dashed line;
($7.8\times 10^{-11}$, $0.80$) --- short-dash-dotted line; ($6.3\times
10^{-11}$, $0.85$) --- long-dash-dotted line.}
\end{figure}

\begin{table}

\caption{Expected event rates in SNU in the small and
large mixing--angle regions for different sets of cross sections (see
text).  The second and third columns are the ratios of the assumed
integrated cross sections for $^8$B and $^7$Be to their mean values
(from Ref.\ \protect{\cite{epv}}); they are all within the range of
uncertainty discussed in the text.  The variation in the cross section of the
CNO--neutrinos is assumed to be 100\% correlated with that of the
$^7$Be neutrinos.}

\begin{tabular}{c c c c c}

     & $^8$B  & $^7$Be & small & large \\ \hline

 1   & 0.5  &  0.2  &  2.5 - 5  &  4 -  5 \\ %\hline
 2   & 0.5  &  1.0  &  3  - 10  &  8 - 14 \\ %\hline
 3   & 0.5  &  1.8  &  4  - 15  & 13 - 21 \\ %\hline
 4   & 1.0  &  0.2  &  5  - 11  &  5 -  8 \\ %\hline
 5   & 1.0  &  1.0  &  5  - 13  & 10 - 16 \\ %\hline
 6   & 1.0  &  1.8  &  5  - 18  & 13 - 22 \\ %\hline
 7   & 1.5  &  0.2  &  6  - 15  &  7 - 11 \\ %\hline
 8   & 1.5  &  1.0  &  7  - 16  & 12 - 18 \\ %\hline
 9   & 1.5  &  1.8  &  6  - 21  & 15 - 25 \\ %\hline\hline
10   & 0.8  &  0.7  &  4.5- 9.5 &  7 - 11 \\ %\hline
11   & 1.2  &  1.3  &  6  - 15  & 13 - 20 \\
\end{tabular}

\end{table}

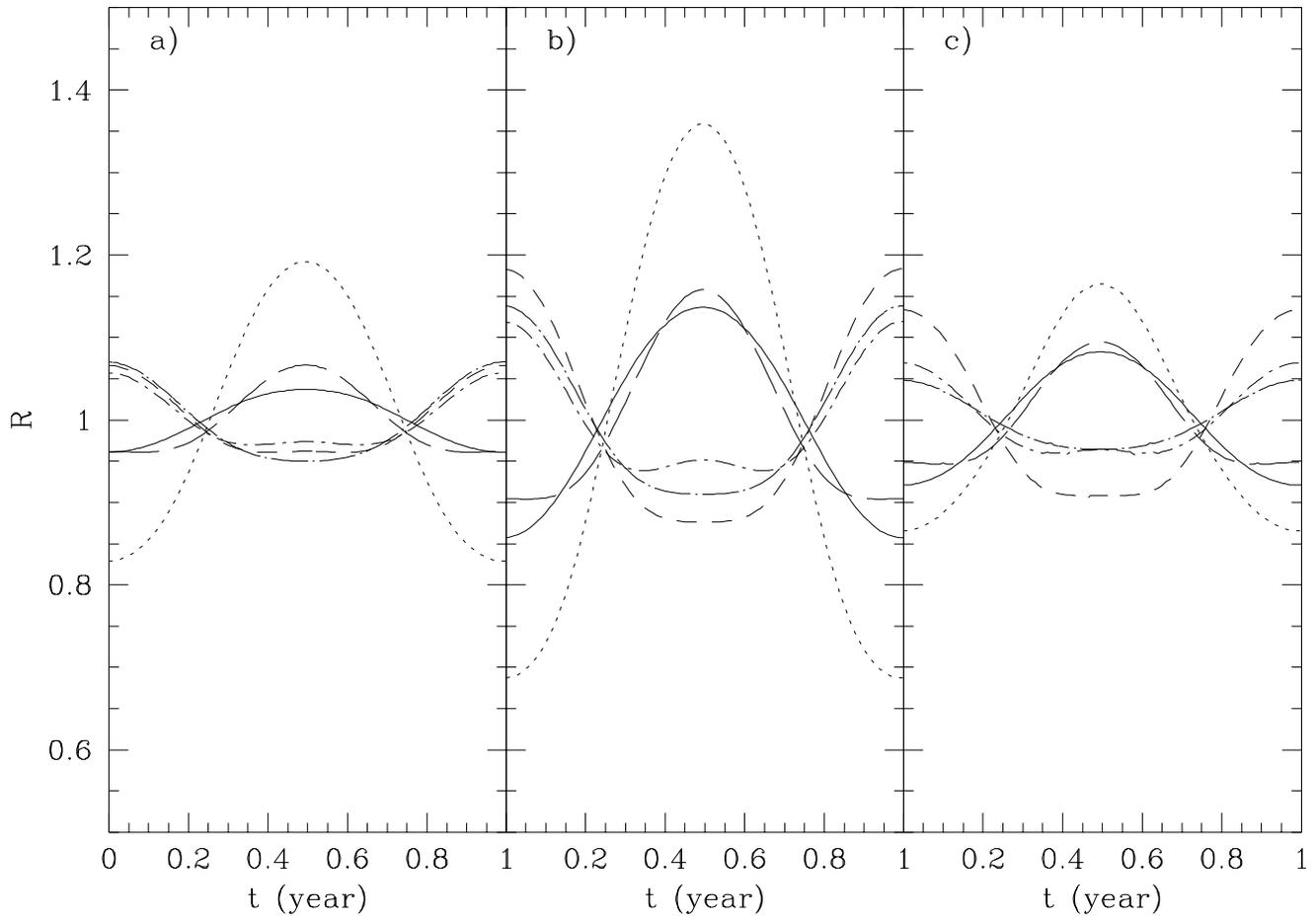
\end{document}